\documentclass[final,5p,times,twocolumn]{elsarticle}

\usepackage{bm} 
\usepackage{graphics}
\usepackage{hyperref}

\usepackage{amssymb}

\newcommand{\bfsfG}{\mbox{\sffamily\bfseries{G}}}
\newcommand{\bfsfI}{\mbox{\sffamily\bfseries{I}}}
\newcommand{\bfsfK}{\mbox{\sffamily\bfseries{K}}}
\newcommand{\bfsfV}{\mbox{\sffamily\bfseries{V}}}

\journal{Photonics and Nanostructures -- Fundamentals and Applications}

\begin{document}

\begin{frontmatter}



\title{Mode expansions in the quantum electrodynamics of photonic media with disorder}



\author{M. Wubs}
\ead{mwubs@fotonik.dtu.dk}
\ead[url]{www.mortensen-lab.org/wubs}
\author{N. A. Mortensen}
\address{Department of Photonics Engineering, Technical University of Denmark, DK-2800 Kgs. Lyngby, Denmark}

\date{\today}

\begin{abstract}
We address two issues in the quantum electrodynamical description of photonic media with some disorder, neglecting material dispersion.
When choosing a gauge in which the static potential vanishes, the normal modes of the medium with disorder satisfy a different transversality condition than the modes of the ideal medium.
Our first result is an integral equation for optical modes such that all perturbation-theory solutions by construction satisfy the desired transversality condition. Secondly, when expanding the vector potential for the medium with disorder in terms of the normal modes of the ideal structure, we find the gauge transformation that conveniently makes the static potential zero, thereby generalizing work by Glauber and Lewenstein [~Phys. Rev. A {\bf 43}, 467 (1991)~]. Our results are  relevant for the quantum optics of disordered photonic crystals.
\end{abstract}


\end{frontmatter}



\section{Introduction}\label{intro}
The quantum optics of random media is a young research field, studying the effects of randomness on quantum correlations and entanglement of quantum states of light in a multi-mode setting~\cite{Patra2000a,Lodahl2005a,Smolka2009a,Ott2010a,Lahini2010a,Cherroret2011a}. Traditionally, random media are studied with randomness against a homogeneous dielectric background. Recently researchers also realized that every real photonic structure, such as a photonic crystal, is in a sense a random medium, since there is inevitably some randomness on top of the ideal dielectric properties~\cite{Koenderink2000a,Koenderink2003a,Kuramochi2005a}. The interplay between the randomness and an ordered inhomogeneous dielectric background can sometimes be exploited. For example, in a photonic-crystal background slow light can promote  localization of light due to even minute random scattering~\cite{John1987a,Patterson2009a,Sapienza2010a,Grgic2010a}.

The typical starting point in the quantum optics of random media is the assumption of a multimode scattering matrix with elements subject to disorder~\cite{Patra2000a,Lodahl2005a,Ott2010a,Lahini2010a}. One level of modeling deeper is the quantum electrodynamics (QED) of these media, to derive the form of the scattering matrix and its dependence on the types of disorder in the medium. Here we aim to contribute to this QED description for spatially inhomogeneous media with some additional disorder. For simplicity we assume that material dispersion can be neglected, as in Refs.~\cite{Knoell1987a,Glauber1991a,Dalton1996a,Dalton1997a,Wubs2003a},  although more general quantized-field theories for dispersive and absorbing inhomogeneous dielectric media have also been developed~\cite{Dung1998a,Suttorp2004a,Bamba2008a,Kheirandish2010a}.

First we will derive a useful new integral equation for the normal optical modes in a photonic medium with disorder. Related work on integral equations and Green-function methods can be found in  Refs.~\cite{Martin1995a,Wubs2002a,Soendergaard2002a,Wubs2004b,Wubs2004a,Rindorf2006a,Guo2007a,Prosentsov2008a,Soerensen2008a,Kristensen2011a}, and on disorder in photonic media in Refs.~\cite{Asatryan1999a,Li2000a,Johnson2002a,Gerace2004a,Hughes2005a,Andreani2006a,LeCamp2007a,Mazoyer2009a,Patterson2010a,Savona2011a}. Instead of an integral equation for the modes involving a disorder potential and the usual Green tensor $\bfsfG$ of the unperturbed medium, we introduce an alternative integral equation involving a kernel $\bfsfK$~\cite{Wubs2004a} that differs from $\bfsfG$ (details below). This $\bfsfK$ emerged naturally in a quantum optical description of light sources and scatterers in a photonic environment~\cite{Wubs2004a}, and has since then been frequently employed in a quantum optics context, e.g. in Refs.~\cite{Hughes2007a,MangaRao2007a,Yao2009a,Yao2010a}. Here instead we propose a novel use to it, in an integral equation that we derive for optical modes of photonic media with disorder. We discuss its specific advantage that arbitrary-order perturbation-theory solutions automatically satisfy a desirable gauge condition.

As our second topic we discuss an alternative to a normal-mode expansion, namely an expansion into modes that get coupled because of a perturbation. Disorder is such a perturbation. Several methods have been developed to describe disorder in photonic crystals~\cite{Asatryan1999a,Li2000a,Hughes2005a,LeCamp2007a,Mazoyer2009a,Patterson2010a,Savona2011a}. These methods are useful both in the classical and in the quantum optics of disordered photonic crystals.  Here our aim  will be to generalize the coupled plane-wave description by Glauber $\&$ Lewenstein~\cite{Glauber1991a}, to a coupled Bloch-mode description for example. Although there will be coupled modes in our theory, it is  different from what is commonly known as a `coupled-mode theory'~\cite{Johnson2002b,Grgic2010a}, which also finds application in quantum optics~\cite{Bromberg2010a}.

The structure of this article is as follows: in Sec.~\ref{normal} we briefly review the quantum electrodynamics of inhomogeneous dielectric media. In Sec.~\ref{gaugenormal} a useful integral equation is derived for the independent optical modes, which is especially well suited for perturbation theory calculations of the effects of disorder on normal modes. Section~\ref{nonnormal} defines the problem to find a convenient gauge transformation when starting from an expansion into modes that are not the normal modes. This transformation is constructed in Sec.~\ref{gaugenonnormalsolved}, specified to the special case of a plane-wave expansion in Sec.~\ref{planewavegauge}, before we conclude in Sec.~\ref{conclusions}.

\section{Normal-mode expansion}\label{normal}
The quantum optical description of the electromagnetic field in a photonic medium with negligible material dispersion starts with the source-free Maxwell equations in matter $\nabla\cdot {\bf D}=0$, $\nabla\cdot {\bf B}=0$, $\dot{\bf D}=\nabla\times {\bf H}$, and $\dot{\bf B}=-\nabla\times {\bf E}$ and the constitutive relations for a lossless nonmagnetic medium,  ${\bf D}({\bf r}) = \varepsilon_{0}\varepsilon({\bf r}){\bf E}({\bf r})$ and ${\bf B}({\bf r}) = \mu_{0} {\bf H}({\bf r})$. We can introduce a vector potential ${\bf A}({\bf r})$ and a scalar potential ${\Phi}({\bf r})$ such that
\begin{eqnarray}
{\bf E} & = & -\nabla\Phi -\dot {\bf A}, \label{EinA} \\
{\bf B} &=& \nabla\times {\bf A}, \label{BinA}
\end{eqnarray}
and where the dot denotes a time derivative. There is gauge freedom, {\em i.e.} one can make combined changes of the vector potential ${\bf A} \rightarrow {\bf A} + \nabla \chi$ and the static potential $\Phi \rightarrow \Phi - \dot \chi$ that leave  the electric and magnetic fields ${\bf E}$ and ${\bf B}$ unaltered. Here $\chi({\bf r},t)$ is an arbitrary scalar function of space and time.

The usual steps from classical to quantum electrodynamics of photonic media are first to choose a convenient  gauge, then to identify the canonical fields, next to express those fields into normal modes, and finally to associate non-commuting operators with them~\cite{Glauber1991a,Dalton1996a,Wubs2003a}. In the first step, the simplest gauge choice for lossless nondispersive photonic media is a generalization of the free-space Coulomb gauge $\nabla \cdot  {\bf A} = 0$, namely the `generalized Coulomb gauge'
\begin{equation}
\nabla \cdot [\,\varepsilon({\bf r}) {\bf A}\,] = 0.
\end{equation}
We call such ${\bf A}$ `generalized transverse' or `$\varepsilon$-transverse', while  Ref.~\cite{Sakoda2005a} uses `quasi-transverse'.
Advantage of this gauge choice is that the corresponding static potential $\Phi$ can be chosen identically zero, which leaves the vector potential as the only canonical variable in the Lagrangian density $\mathcal{L}$  that leads to Maxwell's equations,
\begin{equation}
\mathcal{L} = \frac{1}{2}\varepsilon_{0}\varepsilon({\bf r}) [\dot {\bf A}({\bf r},t)/c]^{2} - \mu_{0}^{-1}[\nabla \times {\bf A}({\bf r},t)]^{2}.
\end{equation}
The field canonically conjugate to the vector potential equals $\varepsilon_{0}\varepsilon({\bf r}) \dot {\bf A}({\bf r}) = - {\bf D}({\bf r})$, and the Hamiltonian density becomes
\begin{equation}\label{Hamiltoniandensity}
\mathcal{H} = \frac{[ {\bf D}({\bf r})]^{2}}{2 \varepsilon_{0}\varepsilon({\bf r})}  + \frac{1}{2\mu_{0}}[\nabla \times {\bf A}({\bf r})]^{2}.
\end{equation}
One can introduce commutation relations for ${\bf A}$ and $-{\bf D}$ directly, but a simpler  equivalent procedure is to expand the fields into normal modes. The wave equation for the vector potential is
\begin{equation}
\nabla \times \nabla\times {\bf A} + \frac{\varepsilon({\bf r})}{c^2}\frac{\partial^{2} {\bf A}}{\partial t^2} =0.
\end{equation}
In terms of the complete set of normal modes ${\bf f}_{\lambda}({\bf r})$ with mode index $\lambda$ that satisfy
\begin{equation}\label{normalmodeseq}
-\nabla \times \nabla\times {\bf f}_{\lambda}({\bf r}) + \varepsilon({\bf r})\frac{\omega_{\lambda}^{2}}{c^2}{\bf f}_{\lambda}({\bf r}) =0,
\end{equation}
the vector potential can be expressed as
\begin{equation}
{\bf A}({\bf r},t) = \sum_{\lambda} \sqrt{\frac{\hbar}{2 \varepsilon_{0}\omega_{\lambda}}}
\left[\,\hat a_{\lambda}(t) {\bf f}_{\lambda}({\bf r})+ \hat a_{\lambda}^{\dag}(t) {\bf f}_{\lambda}^{*}({\bf r}) \,\right].
\end{equation}
This is now a quantum mechanical operator where the creation and annihilation operators $\hat a_{\lambda}^{\dag}$ and $\hat a_{\lambda}$  satisfy the usual harmonic-oscillator commutation relations. Apart from zero-point energies that can be neglected here, the Hamiltonian is a sum of independent harmonic oscillators,
\begin{equation}
H = \sum_{\lambda } \hbar\omega_{\lambda}\,\hat a_{\lambda}^{\dag}\hat a_{\lambda},
\end{equation}
This Hamiltonian defines the modes as `normal modes' and leads to harmonic time dependence for the $\hat a_{\lambda}^{(\dag)}$.   From the latter two equations and the relations~(\ref{EinA},\ref{BinA}), the mode expansions for the electric- and magnetic-field operators follow immediately.

\section{Gauge-respecting perturbation theory for normal modes}\label{gaugenormal}
In order to find the unknown normal modes of a medium $\rm II$ with dielectric function $\varepsilon_{\rm II}({\bf r})$, it is often useful to do this starting from the modes of another medium $\rm I$, for which the normal modes are either known or easier to compute or to interpret. Medium $\rm I$ is often an idealized structure with symmetries that make it easier to classify and find the normal modes, and medium $\rm II$ is its practical realization with some disorder. For example, all real photonic media have some disorder~\cite{Koenderink2000a,Koenderink2003a,Kuramochi2005a}, unwanted or by design~\cite{Barthelemy2008a} or both~\cite{Sapienza2010a}, so that the realized dielectric function $\varepsilon_{\rm II}({\bf r})$ will be the sum of the dielectric function of the ideal structure $\varepsilon_{\rm I}({\bf r})$ plus an inevitable  disorder term $\Delta\varepsilon({\bf r}) = \varepsilon_{\rm II}({\bf r})-\varepsilon_{\rm I}({\bf r})$. The normal modes $\{ {\bf f}_{{\rm II} \mu} \}$ of the medium with disorder differ from the normal modes  $\{ {\bf f}_{{\rm I} \lambda} \}$ of the idealized structure, which for photonic crystals are Bloch modes. For the quantum electrodynamical description this does not pose any formal problems, for {\em in principle} one can follow the quantization procedure as discussed above, and write the vector potential of the medium with disorder as
\begin{equation}\label{AII}
{\bf A}_{\rm II}({\bf r},t) =  \sum_{\mu} \sqrt{\frac{\hbar}{2 \varepsilon_{0} \omega_{\mu}}}
\left[\,\hat a_{{\rm II} \mu}(t) {\bf f}_{{\rm II} \mu}({\bf r})+ \hat a_{{\rm II} \lambda}^{\dag}(t) {\bf f}_{{\rm II} \lambda}^{*}({\bf r}) \,\right],
\end{equation}
The modes $\{ {\bf f}_{{\rm II} \mu} \}$ and the vector potential ${\bf A}_{\rm II}$ satisfy the gauge condition $\nabla \cdot [\,\varepsilon_{\rm II}({\bf r}) {\bf A}_{\rm II}({\bf r})\,]=0$, and this is the gauge in which the static potential for medium $\rm II$ vanishes.

{\em In practice}, it can be quite a challenge to find the normal modes of the vector potential for a complex photonic medium with disorder. Here we develop a calculational tool that can make it simpler. We consider the problem how to find a normal-mode expansion for the vector potential of medium ${\rm II}$, given the normal-mode expansion of ${\bf A}_{{\rm I}}$ of medium $\rm I$ in terms of mode functions ${\bf f}_{{\rm I} \lambda}$ with $\nabla\cdot[\varepsilon_{\rm I}({\bf r}) {\bf f}_{{\rm I} \lambda}({\bf r})]=0$. For medium $\rm II$  we also wish to work in the analogous convenient gauge in which $\nabla\cdot[\varepsilon_{\rm II}({\bf r}) {\bf A}_{{\rm II}}({\bf r})]=0$ so that the static potential vanishes. Our goal is now to find the normal modes ${\bf f}_{{\rm II} \lambda}$. We show that the standard perturbation expansion does not have the desired transversality property, and propose a new  type of perturbation expansion for the normal modes of the vector potential that resolves this issue.

First we derive in a few lines the standard  integral equation for the modes of medium $\rm II$. There is a trivial way to rewrite their defining  equation (\ref{normalmodeseq}) as
 \begin{equation}\label{normalmodesII}
-\nabla \times \nabla\times {\bf f}_{{\rm II} \mu}({\bf r}) + \varepsilon_{\rm I}({\bf r})\frac{\omega_{\mu}^{2}}{c^2}{\bf f}_{{\rm II} \mu}({\bf r}) =  \bfsfV({\bf r},\omega_{\mu})\cdot {\bf f}_{{\rm II} \mu}({\bf r}),
\end{equation}
with the idealized dielectric function $\varepsilon_{\rm I}({\bf r})$ on the left-hand side, and on the right the term with the perturbation potential
 \begin{equation}\label{Vdef}
 \bfsfV({\bf r},\omega)= - [\varepsilon_{\rm II}({\bf r})-\varepsilon_{\rm I}({\bf r})](\omega/c)^{2}\bfsfI = -\Delta\varepsilon({\bf r})(\omega/c)^{2}\bfsfI,
 \end{equation}
where $\bfsfI$ is the unit tensor.
Next we introduce the Green tensor $\bfsfG_{\rm I}$, defined as the solution of the wave equation with delta-function source term,
\begin{equation}\label{Gdefined}
-\nabla \times \nabla\times \bfsfG_{\rm I}({\bf r},{\bf r}',\omega) + \varepsilon_{\rm I}({\bf r})\frac{\omega_{\lambda}^{2}}{c^2}\bfsfG_{\rm I}({\bf r},{\bf r}',\omega) =\delta({\bf r}-{\bf r}')\bfsfI,
\end{equation}
 There is an analogous equation for $\bfsfG_{\rm II}$. By combining the last four equations, we can find a mode ${\bf f}_{{\rm II} \lambda}$  starting with an unperturbed mode ${\bf f}_{{\rm I} \lambda}$, using the exact Lippmann-Schwinger integral equation
 \begin{equation}\label{DSstandard}
 {\bf f}_{{\rm II} \lambda}({\bf r}) =  {\bf f}_{{\rm I} \lambda}({\bf r}) + \int\mbox{d}{\bf r}'\,\bfsfG_{\rm I}({\bf r},{\bf r}',\omega_{\lambda})\cdot \bfsfV({\bf r}',\omega_{\lambda})\cdot {\bf f}_{{\rm II} \lambda}({\bf r}').
 \end{equation}
The ${\bf f}_{{\rm I,II} \lambda}$ have the same label $\lambda$, because the two are related by this integral equation.

We know that $\nabla\cdot[\varepsilon_{\rm I}({\bf r}) {\bf f}_{{\rm I}\lambda}({\bf r})]=0$ and from Eq.~(\ref{normalmodesII}) that  $\nabla\cdot[\varepsilon_{\rm II}({\bf r}) {\bf f}_{{\rm II}\lambda}({\bf r})]$ should vanish for the normal modes of medium $\rm II$, {\em i.e.} for the solution ${\bf f}_{{\rm I,II} \lambda}$ of the integral equation~(\ref{DSstandard}). However, in zero-order perturbation theory one has the solution  ${\bf f}_{{\rm II} \lambda}^{(0)} = {\bf f}_{{\rm I} \lambda}$ that clearly breaks the gauge condition of $\varepsilon_{\rm II}$-transversality. Likewise, in first-order perturbation theory or Born approximation, where one replaces ${\bf f}_{{\rm II} \lambda}$ by ${\bf f}_{{\rm I} \lambda}$ within the integral in Eq.~(\ref{DSstandard}), one finds an improved approximation ${\bf f}_{{\rm II} \lambda}^{(1)}$ that nevertheless breaks the  gauge condition for ${\bf f}_{{\rm II}\lambda}$. The condition is satisfied by the exact (infinite-order) solution only. Now doing perturbation theory up to some finite order of course means that some controlled error is introduced. However, an alternative integral equation that generates solutions that satisfy $\nabla\cdot[\varepsilon_{\rm II}({\bf r}){\bf f}_{{\rm II}\lambda}^{(n)}({\bf r})]=0$ for all orders  of approximation $n$ is clearly to be preferred. Below we present just that.

The Green tensor $\bfsfG_{\rm I}$ of Eq.~(\ref{Gdefined}) can be written as the sum of a generalized transverse part $\bfsfG_{\rm I}^{\rm T}$ and a longitudinal ({\em i.e.} curl-free) part $\bfsfG_{\rm I}^{\rm L}$, which can be expanded in a complete set of generalized transverse modes of medium $\rm I$ as~\cite{Wubs2004a}
\begin{eqnarray}
\bfsfG_{\rm I}^{\rm T}({\bf r},{\bf r}',\omega) & = &  c^{2}\sum_{\lambda}
\frac{{\bf f}_{{\rm I} \lambda}({\bf r}) {\bf f}_{{\rm I} \lambda}^{*}({\bf r}')}{(\omega + \mathrm{i}\eta)^{2} - \omega_{\lambda}^{2}}, \label{GT} \\
\bfsfG_{\rm I}^{\rm L}({\bf r},{\bf r}',\omega) & = & \frac{\delta({\bf r}-{\bf r}')\bfsfI}{\varepsilon_{\rm I}({\bf r}) (\omega/c)^{2}}
- \frac{1}{(\omega/c)^{2}}\sum_{\lambda} {\bf f}_{{\rm I} \lambda}({\bf r}) {\bf f}_{{\rm I} \lambda}^{*}({\bf r}'). \label{GL}
\end{eqnarray}
From this it is easy to see that the total Green function can alternatively be expressed as~\cite{Wubs2004a}
\begin{equation}\label{GinK}
\bfsfG_{\rm I}({\bf r},{\bf r}',\omega) = \bfsfK_{\rm I}({\bf r},{\bf r}',\omega) + \frac{\delta({\bf r}-{\bf r}')\bfsfI}{\varepsilon_{\rm I}({\bf r}) (\omega/c)^{2}},
\end{equation}
where we introduced the tensor $\bfsfK_{\rm I}$ that has the mode expansion
\begin{equation}\label{Kexpansion}
\bfsfK_{\rm I}({\bf r},{\bf r}',\omega) = c^{2}\sum_{\lambda}
\left(\frac{\omega_{\lambda}}{\omega}\right)^{2}\frac{{\bf f}_{{\rm I} \lambda}({\bf r}) {\bf f}_{{\rm I} \lambda}^{*}({\bf r}')}{(\omega + \mathrm{i}\eta)^{2} - \omega_{\lambda}^{2}}.
\end{equation}
An important difference between $\bfsfK$ and $\bfsfG$ is that $\nabla_{\bf r}\cdot[\varepsilon_{\rm I}({\bf r}) \bfsfK_{\rm I}({\bf r},{\bf r'},\omega)]=0$, as follows from the mode expansion  Eq.~(\ref{Kexpansion}), whereas $\nabla_{\bf r}\cdot[\varepsilon_{\rm I}({\bf r}) \bfsfG_{\rm I}({\bf r},{\bf r'},\omega)]\ne 0$. This same Green tensor $\bfsfK$ emerged naturally in the multiple-scattering formalism of light interacting with atoms in photonic media in Ref.~\cite{Wubs2004a}, where there was no way around it, so to say. Here instead we just {\em choose} to rewrite the integral equation~(\ref{DSstandard}) in terms of the new Green tensor, because it produces a more convenient integral equation. After using the identity Eq.~(\ref{GinK}) to make the replacement  of $\bfsfG$ by $\bfsfK$ in the integral equation~(\ref{DSstandard}) and evaluating the integral over the delta-function term of Eq.~(\ref{GinK}) with the help of the expression~(\ref{Vdef}) for $\bfsfV$, we obtain after rearranging the {\em exact} integral equation
\begin{eqnarray}\label{newDSeqforfII}
\varepsilon_{\rm II}({\bf r}){\bf f}_{{\rm II} \lambda}({\bf r}) & =  & \varepsilon_{\rm I}({\bf r}){\bf f}_{{\rm I} \lambda}({\bf r}) \\
& + & \varepsilon_{\rm I}({\bf r})\int\mbox{d}{\bf r}'\,\bfsfK_{\rm I}({\bf r},{\bf r}',\omega_{\lambda})\cdot \bfsfV({\bf r}',\omega_{\lambda})\cdot {\bf f}_{{\rm II} \lambda}({\bf r}'). \nonumber
\end{eqnarray}
Both terms on the right-hand side are divergence-free, the first one by assumption, and for the second term it follows from the $\varepsilon_{\rm I}$-transversality of $\bfsfK_{\rm I}$ as discussed above. It follows that the left-hand side is also divergence-free, so that ${\bf f}_{{\rm II} \lambda}({\bf r})$ on the left is indeed a generalized transverse normal mode for medium~$\rm II$.
Even better, all finite-order perturbation-theory solutions based on this integral equation~(\ref{newDSeqforfII}) have this same property. This is obvious for the zero-order solution ${\bf f}_{{\rm II} \lambda}^{(0)}({\bf r})=\varepsilon_{\rm I}({\bf r}){\bf f}_{{\rm I} \lambda}({\bf r})/\varepsilon_{\rm II}({\bf r})$, found by putting $\bfsfV$ to zero in Eq.~(\ref{newDSeqforfII}). And to first order in the perturbation potential $\bfsfV$, {\em i.e.} in Born approximation, we have
\begin{eqnarray}\label{newDSeqforfIIBorn}
\varepsilon_{\rm II}({\bf r}){\bf f}_{{\rm II} \lambda}^{(1)}({\bf r}) & =  & \varepsilon_{\rm I}({\bf r}){\bf f}_{{\rm I} \lambda}({\bf r}) \\
& + & \varepsilon_{\rm I}({\bf r})\int\mbox{d}{\bf r}'\,\bfsfK_{\rm I}({\bf r},{\bf r}',\omega_{\lambda})\cdot \bfsfV({\bf r}',\omega_{\lambda})\cdot {\bf f}_{{\rm II} \lambda}^{(0)}({\bf r}'). \nonumber
\end{eqnarray}
The left-hand side of this equation is divergence-free, for the same reasons as given above for the left-hand side of Eq.~(\ref{newDSeqforfII}) for the  exact (but implicit) solution.

The new integral equation~(\ref{newDSeqforfII}) is advantageous as compared to Eq.~(\ref{DSstandard}) for another reason. Often one would like to write the solution as a linear combination of solutions of the unperturbed system. In Eq.~(\ref{DSstandard}) one would be tempted to write the ${\bf f}_{{\rm II} \lambda}$ as linear combinations of the ${\bf f}_{{\rm I} \nu}$, but such an expansion would be incomplete because of the different transversality relations of the two types of modes. In Eq.~(\ref{newDSeqforfII}) one can define  ${\bf h}_{{\rm I,II} \lambda}\equiv \varepsilon_{\rm I,II}{\bf f}_{{\rm I,II} \lambda}$, which are divergence-free (just like the transverse plane waves in which they could be expanded). The ${\bf h}_{{\rm II} \lambda}$ can therefore be completely expanded in terms of the ${\bf h}_{{\rm I} \lambda}$, or ${\bf h}_{{\rm II} \lambda}= \sum_{\nu} C_{\lambda\nu}{\bf h}_{{\rm I} \nu}$, and hence the ${\bf f}_{{\rm II} \lambda}$ in terms of the ${\bf f}_{{\rm I} \lambda}$, $\varepsilon_{\rm I}({\bf r})$, and $\varepsilon_{\rm II}({\bf r})$.

To give a simple example of the advantage of our expansion, assume that a smooth background 1D dielectric function $\varepsilon_{\rm B}(z)$ is perturbed and in a finite $z$-interval is replaced by the smooth dielectric function $\varepsilon_{\rm A}(z)$, with discontinuities on the interfaces. 
At an interface, it is well known that the tangential components of the electric field and the normal components of the displacement field are conserved. However, already if we use the zero-order solution of Eq.~(\ref{DSstandard}), the tangential components of the electric field are correctly continuous, but the normal components of the displacement field are not. Conversely, the zero-order solution Eq.~(\ref{newDSeqforfIIBorn}) incorrectly gives discontinuous parallel components of the ${\bf E}$-field but correctly gives continuity of the normal component of the ${\bf D}$ field. So neither zero-order solution is the exact solution of course, but the advantage of the zero-order solution Eq.~(\ref{newDSeqforfIIBorn}) is that it correctly satisfies  the gauge conditions $\nabla\cdot[\varepsilon_{\rm B}(z) {\bf A}({\bf r})]=0$ in the $\rm B$-region and $\nabla\cdot[\varepsilon_{\rm A}(z) {\bf A}({\bf r})]=0$ in the A-region. In this first example we elaborated on the lowest-order approximation. However, the advantage in  quantum electrodynamics of our gauge-respecting series~(\ref{newDSeqforfII}) exists for every order of perturbation theory.

It is also instructive to do an exact calculation with the new integral equation~(\ref{newDSeqforfII}), of local-field effects for example. The electric field in an infinitely small spherical empty cavity around a point-like emitter at ${\bf r}={\bm 0}$ embedded in a homogeneous medium with dielectric function $\varepsilon>1$ is enhanced by a so-called local-field factor $L$, see~\cite{Johnson2005a,Patterson2010a} and the derivation in~\cite{Glauber1991a} based on Eq.~(\ref{DSstandard}), so that the spontaneous-emission rate becomes $L^{2}\sqrt{\varepsilon}$ times the free-space value. Assumed is that the field inside the sphere is constant [${\bf f}({\bf r})={\bf f}({\bm 0})$] and important to know is that only the delta-function contribution $\delta({\bf r}-{\bf r'})\bfsfI/(3 \varepsilon (\omega/c)^{2})$ of $\bfsfG$ contributes to the volume integral in the small sphere, see~\cite{Yaghjian1980a}. Eq.~(\ref{DSstandard}) then immediately gives $L = 1 + (\varepsilon-1)L/(3\varepsilon)$, which has as solution the well-known empty-cavity local-field factor $L = 3 \varepsilon/(2 \varepsilon +1)$.  Likewise one can derive a local-field factor from the new equation~(\ref{newDSeqforfII}), knowing from Eq.~(\ref{GinK}) that the delta-function contribution of $\bfsfK({\bf r},{\bf r'},\omega)$ for the homogeneous medium is $-2\delta({\bf r}-{\bf r'})\bfsfI/(3 \varepsilon (\omega/c)^{2})$: hence the integral equation~(\ref{newDSeqforfII}) gives $L = \varepsilon - \varepsilon (\varepsilon-1)2 L/(3\varepsilon)$. Again we find the $L = 3 \varepsilon/(2 \varepsilon +1)$. This illustrates that the two perturbation series are equivalent, in the sense of producing identical exact (infinite-order) solutions. Nevertheless, finite-order approximations of the equivalent series will generally differ.

\section{Gauge problem for non-normal mode expansions}\label{nonnormal}
For the QED of photonic media with some disorder, either one chooses to expand the field operators in terms of the normal modes $\{ {\bf f}_{{\rm II} \mu} \}$, as above, or one characterizes the disorder by the way the idealized modes $\{ {\bf f}_{{\rm I} \lambda} \}$ become coupled.

We will now turn to the latter approach and expand the vector potential ${\bf A}_{\rm II}$ of medium $\rm II$ in terms of the normal modes $\{ {\bf f}_{{\rm I} \lambda} \}$ of medium $\rm I$. This is certainly possible, but then ${\bf A}_{\rm II}$ satisfies the gauge condition $\nabla \cdot [\,\varepsilon_{\rm I}({\bf r}) {\bf A}_{\rm II}({\bf r})\,]=0$. With that gauge condition for an $\varepsilon_{\rm II}$ medium, a nonzero static potential is required to make the displacement field divergence-free.   Since our goal is to set up a canonical theory in terms of only the vector potentials, a gauge transformation is required. Glauber $\&$ Lewenstein~\cite{Glauber1991a} pointed out this complication of non-normal mode expansions in QED when considering the special case  $\varepsilon_{\rm I}({\bf r})=1$.

We therefore ask whether a gauge transformation exist that transforms  vector and static potentials that satisfy $\nabla \cdot [\,\varepsilon_{\rm I}({\bf r}) {\bf A}({\bf r})\,]=0$ and $\Phi \ne 0$ into  vector and static  potentials satisfying $\nabla \cdot [\,\varepsilon_{\rm II}({\bf r}) {\bf A}({\bf r})\,]=0$ and $\Phi=0$, respectively.  Glauber $\&$ Lewenstein   used properties of the free-space transverse delta function to find the required gauge transformation for the special case  $\varepsilon_{\rm I}({\bf r})=1$~\cite{Glauber1991a}. With applications to disordered photonic crystals in mind, we will now construct a more general gauge transformation for arbitrary smooth positive dielectric functions $\varepsilon_{{\rm I,II}}({\bf r})$. We thereby generalize Ref.~\cite{Glauber1991a} on this point, and our method will be different.

As an {\em Ansatz}, we consider gauge transformations of the form
\begin{equation}\label{ansatzgauge}
\chi({\bf r},t) =  \sum_{\lambda} \sqrt{\frac{\hbar}{2 \varepsilon_{0}\omega_{\lambda}}}\left[ \hat a_{{\rm I} \lambda}(t) \chi_{\lambda}({\bf r})+ {\rm h. c.} \right].
\end{equation}
We wish to show that a particular $\chi({\bf r},t)$ exists such that
the vector potential of the $\varepsilon_{\rm II}$-medium can be expressed as
\begin{equation}\label{AIIexpansionaftergauge}
{\bf A}_{\rm II}({\bf r},t) =  \sum_{\lambda} \sqrt{\frac{\hbar}{2 \varepsilon_{0} \omega_{\lambda}}}
\bigl\{\,\hat a_{{\rm I} \lambda}(t)[ {\bf f}_{{\rm I} \lambda}({\bf r})+ \nabla \chi_{\lambda}({\bf r})]+ {\rm h. c.} \,\bigl\},
\end{equation}
satisfying the generalized Coulomb gauge condition $\nabla\cdot\left[ \varepsilon_{\rm II}({\bf r}){\bf A}_{\rm II}({\bf r},t)\right]=0$. This gives for every separate mode $\lambda$
\begin{equation}\label{gaugeconditionlambda}
\nabla\cdot\bigl\{ \varepsilon_{\rm II}({\bf r}) \left[ {\bf f}_{{\rm I} \lambda}({\bf r})+ \nabla \chi_{\lambda}({\bf r}) \right]\bigl\} =0.
\end{equation}
This  gauge condition defines the `gauge problem' for non-normal modes, and we solve it in the next section.

\section{Gauge transformation for non-normal modes}\label{gaugenonnormalsolved}
In Sec.~\ref{gaugenormal} we expressed the modes of the realized structure $\rm II$ in terms of the modes of the ideal structure $\rm I$. For our purpose of finding the gauge transformation $\chi$, we will consider the less commonly used perturbation expansion in the other direction, expressing the modes of the ideal structure in terms of the modes of the realized structure.  The wave equations for the ideal modes ${\bf f}_{\rm I}$ can be written as
\begin{equation}\label{normalmodesI}
-\nabla \times \nabla\times {\bf f}_{{\rm I} \lambda}({\bf r}) + \varepsilon_{\rm II}({\bf r})\frac{\omega_{\lambda}^{2}}{c^2}{\bf f}_{{\rm I} \lambda}({\bf r}) = - \bfsfV({\bf r},\omega_{\lambda}) \cdot {\bf f}_{{\rm I} \lambda}({\bf r}),
\end{equation}
and notice the minus sign on the right-hand side, as compared to Eq.~(\ref{normalmodesII}). Again an implicit  solution in terms of an integral equation can be given,
 \begin{equation}\label{fIintermsofFII}
 {\bf f}_{{\rm I} \lambda}({\bf r}) =  {\bf f}_{{\rm II} \lambda}({\bf r}) - \int\mbox{d}{\bf r}'\,\bfsfG_{\rm II}({\bf r},{\bf r}',\omega_{\lambda})\cdot \bfsfV({\bf r}',\omega_{\lambda})\cdot {\bf f}_{{\rm I} \lambda}({\bf r}').
 \end{equation}
This exact relation is interesting for our purposes, since we are looking for a gauge transformation of the left-hand side of this equation that makes it $\varepsilon_{\rm II}$-transverse, and the first term on the right-hand side of this equation already has this property. The last term does not, which is consistent with the left-hand side being $\varepsilon_{\rm I}$-transverse.

At this point we use a property of the Green tensor $\bfsfG_{\rm II}$ that we call its generalized Helmholtz decomposition: it has a unique decomposition into an $\varepsilon_{\rm II}$-transverse part  $\bfsfG_{\rm II}^{\rm T}$ and a longitudinal part $\bfsfG_{\rm II}^{\rm L}$, or $\bfsfG_{\rm II} = \bfsfG_{\rm II}^{\rm T} + \bfsfG_{\rm II}^{\rm L}$, with
\begin{equation}
 \nabla\cdot[\varepsilon_{\rm II}({\bf r})\bfsfG_{\rm II}^{\rm T}({\bf r},{\bf r}')]=0, \quad\mbox{and}\quad \nabla\times\bfsfG_{\rm II}^{\rm L}({\bf r},{\bf r}')=0.
\end{equation}
The proof of this property of Green tensors is given in Ref.~\cite{Wubs2004a},  and is based on a similar unique decomposition of vector fields~\cite{Dalton1997a,Wubs2003a}. We can make direct use of this Green-tensor decomposition, simply  by adding a term $\int \bfsfG_{\rm II}^{\rm L}\cdot \bfsfV \cdot {\bf f}_{{\rm I} \lambda}$ on both sides of Eq.~(\ref{fIintermsofFII}). In the same short-hand notation, we obtain
\begin{equation}
 {\bf f}_{{\rm I} \lambda} + \int \bfsfG_{\rm II}^{\rm L}\cdot \bfsfV \cdot {\bf f}_{{\rm I} \lambda} = {\bf f}_{{\rm II} \lambda} - \int\bfsfG_{\rm II}^{\rm T}\cdot \bfsfV\cdot {\bf f}_{{\rm I} \lambda},
\end{equation}
 It is now obvious that the right-hand side is $\varepsilon_{\rm II}$-transverse, so we conclude the same for the left-hand side. Moreover, the second term on the left is longitudinal. Thus we find what we set out to prove, namely that for every mode function ${\bf f}_{{\rm I} \lambda}$, there exists a gauge transformation $\chi_{\lambda}$ that satisfies the gauge condition Eq.~(\ref{gaugeconditionlambda}). But we know more than its mere existence, for we find that the gauge transformation has the particular form
\begin{equation}\label{gaugeterm}
\nabla\chi_{\lambda}({\bf r}) = \int\mbox{d}{\bf r}'\,\bfsfG_{\rm II}^{\rm L}({\bf r},{\bf r}',\omega_{\lambda})\cdot \bfsfV({\bf r}',\omega_{\lambda})\cdot {\bf f}_{{\rm I} \lambda}({\bf r}'),
\end{equation}
where $\bfsfG_{\rm II}^{\rm L}$ can be expressed in terms of the modes ${\bf f}_{{\rm II}\lambda}$, analogous to Eq.~(\ref{GL}).

It is useful to summarize what we have achieved here. Starting with a ${\bf f}_{\rm I}$-mode expansion of the  vector potential (which thereby was $\varepsilon_{\rm I}$-transverse), we have found the gauge transformation~(\ref{gaugeterm}) that makes it $\varepsilon_{\rm II}$-transverse. This is the gauge for which the static potential of the $\varepsilon_{\rm II}$-medium can be chosen identically zero.
Since after the gauge transformation the vector potential is the only canonical field, it is a simple matter to express the electric and magnetic fields, using Eqs.~(\ref{EinA},\ref{BinA}):
\begin{eqnarray}
{\bf E}_{\rm II}({\bf r},t) & = &  - \sum_{\lambda} \sqrt{\frac{\hbar}{2 \varepsilon_{0} \omega_{\lambda}}}
\bigl\{\,\dot a_{{\rm I} \lambda}(t)[ {\bf f}_{{\rm I} \lambda}({\bf r})+ \nabla \chi_{\lambda}({\bf r})]+ {\rm h. c.} \,\bigl\}, \label{EII} \\
{\bf B}_{\rm II}({\bf r},t) & = &  \sum_{\lambda} \sqrt{\frac{\hbar}{2 \varepsilon_{0} \omega_{\lambda}}}
\bigl\{\,\hat a_{{\rm I} \lambda}(t)[ \nabla\times {\bf f}_{{\rm I} \lambda}({\bf r})]+ {\rm h. c.} \,\bigl\}, \label{BII}
\end{eqnarray}
where the gauge term in the magnetic field vanished, being the curl of a gradient. The fact that in this gauge the magnetic field of medium~${\rm II}$ is expanded in the normal mode functions of medium~${\rm I}$ can be numerically advantageous. The electric field has an additional non-vanishing gauge term, taking over the role of  - and mathematically identical to - the static potential in the initial gauge. The displacement field ${\bf D}_{\rm II}$ is given by $\varepsilon_{0}\varepsilon_{\rm II}({\bf r}){\bf E}_{\rm II}({\bf r})$,  and hence is divergence-free as it should.

\section{Special case: plane waves as non-normal modes}\label{planewavegauge}
 Here we study how our results of the previous section simplify in the special case that $\varepsilon_{\rm I}({\bf r})=1$, in other words if for the ideal disorder-free medium $\rm I $ we choose free space, with its transverse plane-wave modes $\exp({\mathrm i}{\bf k}\cdot{\bf r}){\bf e}_{{\bf k}\sigma}$, where $\sigma=1,2$ labels the two orthogonal unit vectors ${\bf e}_{{\bf k}\sigma}$ perpendicular to the wave vector ${\bf k}$. In medium $\rm II$ with dielectric function $\varepsilon_{\rm II}({\bf r})$, these transverse plane waves are not the normal modes of the vector potential ${\bf A}_{\rm II}$. But we can just expand ${\bf A}_{\rm II}$ in transverse plane waves, and use the gauge transformation $\chi$ that we found in the previous section to end up in the gauge in which the static potential vanishes.
Using the explicit form of the gauge term~(\ref{gaugeterm}) and the expansion~(\ref{GL}) of the longitudinal Green tensor $\bfsfG_{\rm II}$ into the normal modes ${\bf f}_{{\rm II} {\bf k}\sigma}$, we find
\begin{eqnarray}
&& \nabla \chi_{{\bf k}\sigma}({\bf r})  =  \left[\frac{1-\varepsilon_{\rm II}({\bf r})}{\varepsilon_{\rm II}({\bf r})}\right]e^{{\mathrm i}{\bf k}\cdot{\bf r}}{\bf e}_{{\bf k}\sigma} \\
 & + & \sum_{{\bf k}\sigma}{\bf f}_{{\rm II} {\bf k}\sigma}({\bf r})\int\mbox{d}{\bf r}'\,{\bf f}_{{\rm II} {\bf k}\sigma}({\bf r'})\cdot
\left[1-\varepsilon_{\rm II}({\bf r}')\right]e^{{\mathrm i}{\bf k}\cdot{\bf r}'}{\bf e}_{{\bf k}\sigma}. \nonumber
\end{eqnarray}
Therefore, the spatial dependence $[ {\bf f}_{{\rm I} \lambda}({\bf r})+ \nabla \chi_{\lambda}({\bf r})]$ of the vector potential ${\bf A}_{\rm II}$ in Eq.~(\ref{AIIexpansionaftergauge}) in our special case becomes
\begin{equation}\label{spatialdependence}
 \frac{e^{{\mathrm i}{\bf k}\cdot{\bf r}}{\bf e}_{{\bf k}\sigma}}{\varepsilon_{\rm II}({\bf r})}
  + \sum_{{\bf k}\sigma} {\bf f}_{{\rm II} {\bf k}\sigma}({\bf r})\int\mbox{d}{\bf r}'\,{\bf f}_{{\rm II} {\bf k}\sigma}({\bf r'})\cdot
\left[1-\varepsilon_{\rm II}({\bf r}')\right]e^{{\mathrm i}{\bf k}\cdot{\bf r}'}{\bf e}_{{\bf k}\sigma}. \nonumber
\end{equation}
It follows that the vector potential indeed satisfies  $\nabla\cdot\left[\varepsilon_{\rm II}({\bf r}){\bf A}_{\rm II}({\bf r})\right]=0$, that the electric field ${\bf E}_{\rm II}= -\dot {\bf A}_{\rm II}$ satisfies the same condition, and  that ${\bf D}_{\rm II} = \varepsilon_{0}\varepsilon_{\rm II}({\bf r}){\bf E}_{\rm II}$ is divergence-free.

For the magnetic field ${\bf B}_{\rm II}$ it follows from Eq.~(\ref{BII}) that it can be fully expanded in the free-space transverse plane waves. The displacement field has the same spatial dependence as ${\bf A}_{\rm II}$ in Eq.~(\ref{spatialdependence}), but multiplied by $\varepsilon_{\rm II}({\bf r})$. The result is a transverse plane wave, plus the divergence-free gauge term that is not a plane wave. By Fourier-expanding the gauge term into transverse plane waves and regrouping, one could alternatively expand the displacement field into transverse plane-wave modes, which is the approach of Glauber $\&$ Lewenstein~\cite{Glauber1991a}.

\section{Conclusions}\label{conclusions}
After a short review of the quantization of the electromagnetic field in inhomogeneous dispersionless dielectrics, we argued that a convenient choice of gauge is the one for which the static potential vanishes, leaving the vector potential as the only dynamical field. In such a gauge, a mode function of the vector potential for an idealized structure has different transversality properties than for a realistic structure, due to disorder.

Perturbative solutions based on a standard integral equation for modes in disordered photonic media do not give the desired transversality property. Here we proposed an improved integral equation, in terms of the Green tensor $\bfsfK$ \cite{Wubs2004a} rather than $\bfsfG$, that does give such solutions to any order of perturbation.  This can become a useful numerical tool to find exact or approximate normal modes, for example in photonic crystals with disorder.

Besides normal-mode expansions, one can alternatively express the field operators in terms of modes that are not independent, with couplings amongst them due to the disorder. Starting with an expansion of the vector potential of a disordered medium in terms of modes of an idealized structure, we constructed the gauge transformation that makes the static potential identically zero, and obtained expansions for the field operators into modes of the idealized structure plus gauge terms.

We focused on issues  related to the choice of a gauge in the quantum electrodynamics of photonic media. Our results will be useful for developing a full QED theory of modes in complex photonic systems that become coupled due to disorder.

\section*{Acknowledgments}
\noindent Discussions with A. Lagendijk and J. R. Ott are gratefully acknowledged. M. W. acknowledges financial support by The Danish Research Council for Technology and Production Sciences  (FTP grant $\#274-07-0080$), and by the Otto M{\o}nsted Foundation.

\appendix





\bibliographystyle{elsarticle-num}
\bibliography{<your-bib-database>}







\end{document}